\documentclass[lettersize,journal]{IEEEtran}
\usepackage{amsmath,amsfonts}
\usepackage{algorithmic}
\usepackage{algorithm}
\usepackage{array}
\usepackage[caption=false,font=normalsize,labelfont=sf,textfont=sf]{subfig}
\usepackage{textcomp}
\usepackage{stfloats}
\usepackage{url}
\usepackage{verbatim}
\usepackage{graphicx}
\usepackage{cite}
\usepackage{bm}
\usepackage{multirow}
\usepackage[normalem]{ulem}
\useunder{\uline}{\ul}{}
\usepackage{tipa}
\usepackage{diagbox}
\usepackage{footmisc}
\usepackage{hyperref}
\usepackage[hyphenbreaks]{breakurl}
\usepackage{xcolor}
\usepackage{algorithm}
\usepackage{algorithmic}
\usepackage{multirow}
\usepackage{bbding}
\usepackage{booktabs}
\usepackage{amssymb,amsmath,bm}

\begin{document}

\title{ERVQ: Enhanced Residual Vector Quantization with Intra-and-Inter-Codebook Optimization for Neural Audio Codecs}

\author{Rui-Chen Zheng, ~Hui-Peng Du, ~Xiao-Hang Jiang, ~Yang Ai,~\IEEEmembership{Member,~IEEE},~Zhen-Hua Ling,~\IEEEmembership{Senior Member,~IEEE}

\thanks{This work was funded by the National Nature Science Foundation of China under Grant 62301521 and the Anhui Provincial Natural Science Foundation under Grant 2308085QF200. (Corresponding author: Yang Ai)}
\thanks{Rui-Chen Zheng, Hui-Peng Du, Xiao-Hang Jiang, Yang Ai and Zhen-Hua Ling are with the National Engineering Research Center of Speech and Language Information Processing, University of Science and Technology of China, Hefei, 230027, China (e-mail: zhengruichen@mail.ustc.edu.cn, redmist@mail.ustc.edu.cn, jiang\_xiaohang@mail.ustc.edu.cn, yangai@ustc.edu.cn, zhling@ustc.edu.cn).}
}



\maketitle

\begin{abstract}
Current neural audio codecs typically use residual vector quantization (RVQ) to discretize audio signals. However, they often experience codebook collapse, which reduces the effective codebook size and leads to suboptimal performance.
To address this problem, we propose \textbf{E}nhanced \textbf{R}esidual \textbf{V}ector \textbf{Q}uantization (\textbf{ERVQ}), a novel enhancement strategy for the RVQ framework in neural audio codecs. ERVQ mitigates codebook collapse and boosts codec performance through both intra- and inter-codebook optimization. 
Intra-codebook optimization incorporates an online clustering strategy and a code balancing loss to ensure balanced and efficient codebook utilization. Inter-codebook optimization improves the diversity of quantized features by minimizing the similarity between successive quantizations. 
Our experiments show that ERVQ significantly enhances audio codec performance across different models, sampling rates, and bitrates, achieving superior quality and generalization capabilities. It also achieves 100\% codebook utilization on one of the most advanced neural audio codecs.
Further experiments indicate that audio codecs improved by the ERVQ strategy can improve unified speech-and-text large language models (LLMs). Specifically, there is a notable improvement in the naturalness of generated speech in downstream zero-shot text-to-speech tasks. 
Audio samples are available on the project page\footnote{\url{https://zhengrachel.github.io/ERVQ/}.}.

\end{abstract}


\section{Introduction}
\label{Section: Introduction}

\IEEEPARstart{A}{udio} codec, an important signal processing technique, compresses audio signals into discrete codes and reconstructs the original audio from these codes. It plays a pivotal role in fields such as audio communication and transmission \cite{brandenburg1994iso, kroon1986regular, salami1994toll}. Recently, audio codecs have also found applications in various downstream tasks, such as unified speech-and-text large language models (LLMs) \cite{borsos2023audiolm, wang2023neural, shen2023naturalspeech, chen2023lauragpt, junaturalspeech}, where the discrete codes generated by the audio codec serve as intermediate representations for speech generation purpose.

The primary objective of an audio codec is to reduce the amount of data required to store or transmit audio signals without significantly degrading the sound quality. A closely related idea is the use of vector quantization (VQ) \cite{van2017neural} as an information bottleneck to eliminate redundant or irrelevant information from the audio signal, thereby achieving data compression \cite{garbacea2019low, chen2021tenc, vali21_interspeech, lee2022progressive}. 
A typical neural network-based audio codec often adopts an encoder-decoder framework. The encoder first compresses the waveform into a compact and deep representation. Then, a VQ block is employed to quantize the intermediate features. Finally, the decoder reconstructs the waveform from the quantized representation. 
A recent audio codec model, SoundStream \cite{zeghidour2021soundstream}, has further advanced this approach by introducing residual VQ (RVQ), which has demonstrated impressive performance. RVQ recursively quantizes residuals using multiple VQ codebooks to represent intermediate features after the initial quantization step, which has now become a core technology in modern neural audio codecs \cite{wu2023audiodec, jenrungrot2023lmcodec, siuzdak2024snac, zhang2024high, xu2024lightcodec, huang-etal-2024-repcodec}. 

Although RVQ is a popular method for training audio codecs, it faces several challenges in practical applications. The most severe issue is codebook collapse, where only a tiny fraction of code vectors (i.e., entries in the codebook) for each VQ codebook in RVQ receive useful gradients for optimization. Consequently, most code vectors become invalidated, remaining neither updated nor used, leading to codebook under-utilization \cite{dhariwal2020jukebox, kumar2024high}. This reduction in the effective codebook size results in an implicit decrease in the target bitrate, thereby limiting the ability of audio codecs to learn large codebooks and resulting in suboptimal reconstruction quality. Previous research on audio codecs has suggested some strategies to address the codebook collapse issue, such as appropriate codebook initialization and re-initialization \cite{dhariwal2020jukebox, zeghidour2021soundstream, defossez2023high}. However, these strategies only optimize each VQ module within the RVQ framework individually, neglecting the intrinsic characteristics of RVQ as a residual structure. Consequently, they do not adequately resolve the codebook collapse issue. 

To address the codebook collapse issue and further improve the quantization capabilities of audio codecs, in this paper, we propose a novel improvement strategy, named \textbf{ERVQ}, \textbf{E}nhanced \textbf{R}esidual \textbf{V}ector \textbf{Q}uantization for neural audio codecs through intra- and inter-codebook optimization.  
Unlike previous methods that only optimize each VQ independently, the proposed ERVQ also takes into account the structural characteristics of RVQ, offering a holistic optimization approach. 
ERVQ enhances audio codec performance through both intra- and inter-codebook optimization techniques. 
For intra-codebook optimization, we adopt an online clustering strategy and introduce a code balancing loss to tackle the codebook collapse problem and improve the expressiveness of each VQ. 
Specifically, the online clustering strategy calculates the average codebook usage at each iteration and dynamically reinitializes the codebook based on this usage, ensuring each code vector is optimized. The code balancing loss assumes a uniform prior distribution of codes and improves codebook utilization by minimizing the difference between the prior and posterior code distributions. 
For inter-codebook optimization, we propose maximizing the difference in the content quantized by each adjacent VQ to reduce information redundancy between quantized features and further boost the overall expressiveness of RVQ, which is achieved by minimizing a sum of the structural similarity (SSIM) \cite{wang2004image} loss. 

ERVQ can be easily implemented in various neural audio codecs. 
Our experiments on neural audio codecs with various structures at different sampling rates and bitrates demonstrate that the ERVQ strategy consistently improves performance, showcasing its effectiveness and generalization ability.  Since ERVQ is designed for the training stage, it does not affect the inference efficiency of the original model. 
In a codebook analysis experiment, ERVQ improves the codebook utilization rate from 14.7\%$\sim$41.2\% to 100\% for APCodec \cite{ai2024apcodec}.
Furthermore, we have tested an ERVQ-enhanced audio codec \cite{du2024funcodec} combined with a unified speech-and-text LLM \cite{chen2023lauragpt} on downstream zero-shot text-to-speech (TTS) tasks. The results exhibit that the speech synthesized with tokens from the ERVQ-enhanced audio codec demonstrates superior naturalness and stronger expressiveness over baselines. This improvement benefits from the diversity of code vectors by the ERVQ strategy, leading to varied and adaptable speech styles, and further highlighting the powerful codebook learning capability of ERVQ.

Our contribution can be summarized as follows: 
\begin{itemize}
    \item We address the critical issue of codebook collapse in the RVQ framework of neural audio codecs, which significantly limits performance due to the underutilization of effective codebook capacity. To this end, we propose ERVQ, an innovative strategy comprising intra- and inter-codebook optimizations. The intra-codebook optimization includes an online clustering strategy for dynamically reinitializing unused codewords and a code balancing loss to improve utilization rates. Meanwhile, the inter-codebook optimization introduces a structural similarity loss to minimize similarity across adjacent quantizers, enhancing overall expressiveness. 
    \item Comprehensive experiments conducted across various audio codecs with diverse architectures and bitrates validate the effectiveness and generalizability of ERVQ, demonstrating substantial improvements in audio quality. 
    \item Additionally, we integrate the ERVQ-enhanced FunCodec into the LauraGPT model, illustrating its ability to enhance downstream zero-shot text-to-speech tasks. 
    \item The ERVQ strategy requires no additional parameters, is straightforward to implement, and can seamlessly integrate into existing audio codec architectures, underscoring its practicality and wide applicability. 
\end{itemize}

The rest of the paper is organized as follows: we briefly review of the RVQ structure and recent advancements in neural audio codecs, along with their RVQ optimization strategies.  Next, we provide details of our proposed ERVQ strategy in Section \ref{Section: Proposed Methods}. We then present our experimental results in Section \ref{sec: Experiments: Audio Coding} and \ref{sec: Experiments: LLM Applications}, and finally, we draw conclusions in Section \ref{sec: Conclusion}.

\section{Related Work}
\label{Section: Related Work}

This paper focuses on enhancing the quantization capability of the RVQ framework to strengthen the performance of neural audio codecs. In this section, we first review the specific structure of the RVQ framework, detailing its hierarchical quantization process and the role of residual vector quantization in efficiently encoding audio features. We then provide an overview of representative works in recent neural audio codecs adopting RVQ structure and their strategies for improving the RVQ module.

\subsection{RVQ Module}
Neural audio codecs have witnessed significant advancements with the introduction of RVQ as a central mechanism for efficient and high-quality audio compression.  RVQ was first introduced by the pioneering model SoundStream \cite{zeghidour2021soundstream}, which implemented it with a fully causal convolutional encoder-decoder network, achieving low-latency, high-quality audio compression. 

Assume that the RVQ structure contains $M$ VQs, each associated with a trainable codebook $\mathbf{C}^m\in \mathbb R^{K\times N}$, where $m=1,2,\dots,M$, and $K$ and $N$ represent the size and dimension of the codebook, respectively. The quantization process of RVQ operates as follows: For the first VQ, its input consists of the continuous latent features $\mathbf{Z} \in \mathbb{R}^{T\times F}$ generated by the encoder, where $T$ and $F$ denote the number of frames and the feature dimensions, respectively.
In most cases, $N=F$. In some architectures such as factorized codes \cite{kumar2024high}, a projection is applied to reduce the dimensionality of the feature space before quantization, leading to $N < F$. 
For the $i$-th frame $\mathbf{z}_i$ of $\mathbf{Z}$, the Euclidean distance between $\mathbf{z}_i$ and each code vector in  $\mathbf{C}^1$ is calculated. The codec vector with the smallest distance is then selected as the quantized feature $\mathbf{\hat{z}}^1_i$, and its corresponding index is saved as the quantized codes $c_i^1$. Next, the quantization residual $\mathbf{\tilde{z}}_i^1=\mathbf{z}_i-\mathbf{\hat{z}}^1_i$ is computed as the input for the second VQ. This process is repeated sequentially for all $M$ VQs. The final quantized feature is computed as the sum of the outputs from all VQs,  i.e. $\mathbf{\hat{z}}_i=\sum_{m=1}^M \mathbf{\hat{z}}^m_i$. The correpsonding indices $c_i^1, c_i^2, \dots, c_i^M$ serve as the quantized discrete tokens for the $i$-th frame of the continuous latent features input to the RVQ. The functionality of the RVQ structure can be expressed as:
\begin{equation}
    \mathbf{\hat{z}}_i, (c_i^1, c_i^2, \dots, c_i^M) = RVQ(\mathbf{z}_i|\mathbf{C}^1,\dots,\mathbf{C}^M).
\end{equation}
During the training of a neural audio codec, the codebook of each quantizer in the RVQ structure is optimized using the training methods of VQ-VAE\cite{van2017neural}, which include a codebook loss and a commitment loss, or through exponential moving average updates proposed in VQ-VAE-2 \cite{razavi2019generating }.

\subsection{Recent Neural Audio Codecs with RVQ Structures}
Since the introduction of the RVQ structure for quantization in audio codecs by SoundStream\cite{zeghidour2021soundstream}, the RVQ module has emerged as a widely adopted component in neural audio codec architectures. 
Encodec \cite{defossez2023high} closely followed the SoundStream \cite{zeghidour2021soundstream} recipe, utilizing a multi-scale Short-Time Fourier Transform (STFT) discriminator with a multi-scale spectral reconstruction loss, resulting in improved audio quality and better reconstruction of fine details. To mitigate the issue of codebook collapse in RVQ, both models initialized code vectors using k-means clustering to avoid low codebook usage due to poor initialization. Additionally, they randomly reset unused code vectors over several batches in each individual VQ module following \cite{dhariwal2020jukebox}. 
Building on these efforts, DAC \cite{kumar2024high} introduced an adaptation of the improved VQGAN image model \cite{yu2021vector} to the domain of high-fidelity audio compression with multi-scale STFT discriminators and a multi-scale mel-reconstruction loss. It employed two key techniques from the improved VQGAN image model \cite{yu2021vector} to improve codebook usage: factorized and L2 normalized codes. Factorization decouples code lookup from the code vector, using only the principal components of the input vector for code lookup in a low-dimensional space, while the code vector resides in a high-dimensional space. L2 normalization converts Euclidean distance to cosine similarity, enhancing stability and quality.

In addition to standard RVQ designs, several VQ variants have been proposed to improve quantization flexibility. HiFi-Codec \cite{yang2023hifi} observed that the initial codebook within RVQ often stores the majority of information, leaving subsequent codebooks to capture only finer details. To distribute the quantization burden more effectively, HiFi-Codec introduced Group RVQ (GRVQ), a variation of RVQ. GRVQ splits the vector to be quantized into multiple groups, applies RVQ independently to each group, and combines the results for waveform reconstruction. This grouping strategy reduces the quantization difficulty for the first codebook and improves overall performance. Moreover, SNAC \cite{siuzdak2024snac} applies multiscale RVQ for adaptive bitrate control. Other works have proposed architectural modifications to improve quantization effectiveness. A notable example is HARP-Net \cite{petermann2021harp}, which introduces skip autoencoders between each encoder-decoder pair. Each skip connection includes its own quantization module, enabling scalable bitrate control by aggregating codes from multiple levels of representation.
Additional methods focus on improving quantization dynamics directly. CBRC \cite{xu23_interspeech} employs beam search to enhance quantization accuracy through multi-path selection, while CSVQ \cite{jiang22_interspeech} fuses features across encoder and decoder scales to enrich representation diversity and reduce redundancy.
 
Despite the innovations in these methods, several challenges remain unresolved. For instance, the initialization and random reset strategies in SoundStream and Encodec provide limited improvements in codebook utilization, leaving room for enhancing the overall audio codec quality. Factorized codes, as introduced in DAC, require a fully connected network to map quantized vectors to a low-dimensional space, introducing additional parameters to the codec. Similarly, GRVQ modifies the model structure, potentially increasing parameter count and altering the target bitrate, which may complicate its integration into existing codecs. Most notably, these strategies focus primarily on optimizing each VQ module independently, overlooking the hierarchical and residual structure inherent to RVQ.

Additionally, prior works have proposed entropy-based regularization strategies to improve representation compactness or control bitrate in neural speech coding. For example, \cite{kankanahalli2018end, kleijn2018wavenet, zhen19_interspeech, yang2021source, jiang2022end} introduce entropy loss terms in the context of scalar or waveform-level quantization, often in combination with entropy coding schemes such as Huffman or arithmetic coding. However, these methods primarily aim to match target bitrates or compress output distributions, and are not directly designed to address codebook collapse in residual vector quantization.

To address these limitations, we propose ERVQ as a novel improvement strategy for the RVQ framework in audio codec models. While enhancing each VQ independently through intra-codebook optimization, the proposed ERVQ also considers the structural characteristics of RVQ for an inter-codebook optimization approach.
ERVQ is designed for simplicity and generality, requiring minimal implementation effort. It can be easily implemented in various neural audio codecs. 
This proposed strategy neither introduces additional parameters to the model nor requires any changes to the model structure. It can be combined with most existing RVQ improvement strategies, thus offering strong generalization capabilities.  

\section{Proposed Methods}
\label{Section: Proposed Methods}
\begin{figure*}[t]
\centering
\includegraphics[width=0.95\textwidth]{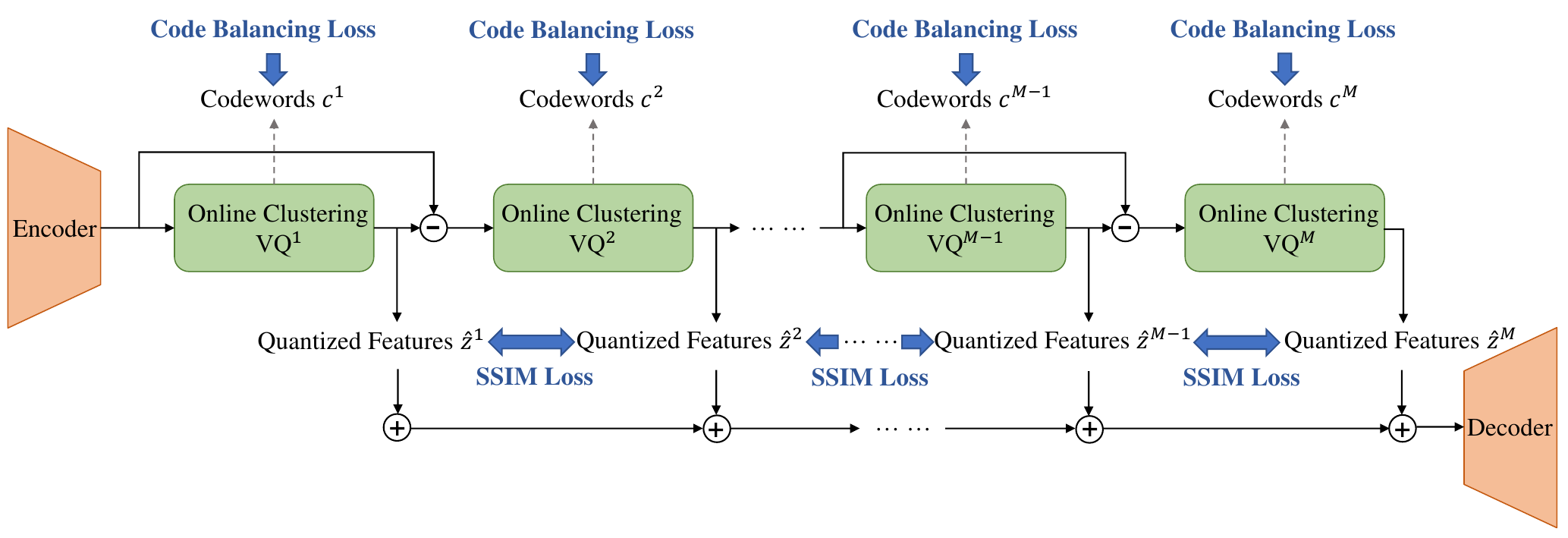} 
\caption{Details of the proposed ERVQ strategy.}
\label{fig: ERVQ}
\end{figure*}

To tackle the codebook collapse issue in RVQ and enhance its quantization capability, the proposed ERVQ strategy improves RVQ through both intra- and inter-codebook optimization. The details of the ERVQ strategy are illustrated in Figure \ref{fig: ERVQ} and described below. 

\subsection{Intra-Codebook Optimization}
Intra-codebook optimization primarily addresses the issue of codebook collapse by ensuring that each code vector within the codebook is selected and optimized during training. To achieve this, we employ an online clustering strategy for each codebook and incorporate an additional code balancing loss to regulate the selection of codes. 

\subsubsection{Online Clustering Strategy}
\label{Section: Online Clustering Strategy}
Inspired by \cite{zheng2023online}, the online clustering strategy builds a dynamically initialized codebook for each VQ in RVQ. Unlike Jukebox \cite{dhariwal2020jukebox} which reinitializes code vectors unused for several batches with randomly sampled encoding features, online clustering strategy aims to modify less-used or unused code vectors more frequently than those regularly used in each batch.
To achieve this, we first accumulatively calculate the average usage $U_k^{(t)}$ of the $k$-th code vector $\textbf{e}_k^{(t)}$ within the codebook in the $t$-th training batch:
\begin{equation}
    U_k^{(t)} = U_k^{(t-1)} \cdot \gamma + \frac{u_k^{(t)}}{L} \cdot(1-\gamma),
\end{equation}
where $u_k^{(t)}$ is the number of frame-level encoded features quantized to the code vector $\textbf{e}_k^{(t)}$ in the $t$-th training batch, and $L$ denotes the total number of frame-level encoded features to be quantized in a batch. $\gamma$ is a decay hyperparameter with a value in $(0, 1)$.  $U_k^{(0)}$ is initialized as zero. 
Next, a decay value $d_k^{(t)}$ for each code vector within the codebook is calculated using the accumulative average usage $U_k^{(t)}$:
\begin{equation}
    d_k^{(t)} = \exp ^{-U_k^{(t)}K\frac{10}{1-\gamma}-\epsilon},
\end{equation}
where $K$ is the codebook size, and $k \in \{1,2,\dots, K\}$. The term $\frac{1}{1-\gamma}$ compensates for the limited update dynamics of the moving average in Equation (2) when $\gamma$ is close to 1, as is typical in practice. The scaling factor 10 is adopted from \cite{zheng2023online} and controls the overall decay strength. 
This formulation allows the decay coefficient $d_k^{(t)}$ to remain sensitive to differences in codeword activity even under strong momentum smoothing.
Additionally, the small constant $\epsilon$ is added to the exponent as a safeguard against excessively fast decay of $d_k^{(t)}$ for highly active codewords, ensuring a minimal but non-zero update rate for them. 
The inactive code vectors can then be reinitialized as follows:
\begin{equation}
   \textbf{e}_k^{(t)} = \textbf{e}_k^{(t-1)}\cdot(1-d_k^{(t)})+\hat{\textbf{z}}_k^{(t)}\cdot d_k^{(t)},
\end{equation}
where $\hat{\textbf{z}}_k^{(t)}$ is the selected encoded features (i.e., anchors). 
Note that we do not define a codeword as “inactive” using any fixed threshold. Instead, the decay coefficient $d_k^{(t)}$ varies continuously based on usage frequency $U_k^{(t)}$. Codewords used less frequently are assigned higher decay values and updated more aggressively, while frequently used ones are adjusted more conservatively. This soft prioritization avoids the instability of manual thresholds and ensures dynamic rebalancing during training.

To update underutilized codewords, we empirically employ a probabilistic random sampling method to sample anchors from encoded features $\textbf{z}_i^{(t)}$ according to the following probability distribution:
\begin{equation}
   \textbf{P}_{Prob}=[Softmax(D_{1,k}), \dots, Softmax(D_{L,k})]^\top,
\end{equation}
where $D_{i,k} = ||\textbf{z}_i^{(t)}-\textbf{e}_k^{(t)}||^2$ is the distance between the code vectors and the encoded features. One feature is then sampled from according to the distribution $\textbf{P}_{Prob}$ and used as the anchor $\hat{\textbf{z}}_k^{(t)}$ to update the codeword as defined in Equation (4).

\subsubsection{Code balancing loss}
Using the online clustering strategy can activate unused codewords and ensure their utilization rate to be greater than zero. However,  it cannot guarantee the extent of their usage. A potential issue is that some codewords may be used at a so low frequency that it is almost negligible. Therefore, we introduce a code balancing loss during the backpropagation process for each VQ. Specifically, assume a prior uniform discrete distribution for the codes selected for quantization as follows:
\begin{equation}
    \textbf{P}_{prior} = \left[\frac{1}{K}, \frac{1}{K}, \dots, \frac{1}{K}\right]^\top \in\mathbb{R}^{K}.
\end{equation}
This prior distribution encourages all codebook embeddings to be uniformly used, thereby maximizing their information capacity according to the principle of maximum entropy. The posterior code distribution $\textbf{P}_{post}^m$ of $m$-th VQ is approximated by the frequency with which each code is chosen during training. 
To elaborate further, suppose a training batch contains $B\times L$ features to be quantized, where each feature's quantization result is represented as a one-hot vector $\textbf{p}_i$ of length $K$. The posterior distribution $\textbf{P}_{post}^m$ is then approximated as the average of all these one-hot vectors within the batch:
\begin{equation}
   \textbf{P}_{post}^m = \frac{1}{B\times L}\sum_{i=1}^{B\times L}\textbf{p}_i=[f_1^m, f_2^m, \dots, f^m_K]^\top,
\end{equation}
where $f_k^m$ represents the frequency of which the $i$-th code is chosen during training for $m$-th VQ. 
Although code selection involves an argmax operation, we follow common practice in VQ models and adopt the straight-through estimator \cite{van2017neural}, which enables gradient flow through the quantization step.
The formulation in Equation (7) also ensures differentiability as the gradients propagate through the one-hot representations during backpropagation.
To effectively avoid codebook collapse, the discrepancy between the prior and posterior distributions is minimized using the following code balancing loss function:
\begin{equation}
   \begin{aligned}
    \mathcal{L}_{balancing} &= \sum\nolimits_{m=1}^M CrossEntropy(\textbf{P}_{post}, \textbf{P}_{prior}) \\
    &=-\sum\nolimits_{m=1}^M \sum\nolimits_{k=1}^Kf_k^m\log {\frac{1}{K}}.
    \end{aligned}
\end{equation}
This loss remains non-trivial during training because $\textbf{P}_{post}^m$ is computed using differentiable approximations of one-hot vectors derived from the quantizer outputs. As a result, the cross-entropy loss yields meaningful gradients, guiding the model to promote more uniform codebook usage without being constant with respect to model parameters. 

While conceptually related to prior entropy-based regularization techniques \cite{kankanahalli2018end, kleijn2018wavenet, jiang2022end, zhen19_interspeech, yang2021source}, our loss operates differently: it is not tied to bitrate control or entropy coding, but rather focuses on maintaining active and balanced codeword usage across VQ stages to avoid collapse in RVQ-based architectures.
This method also differs from the strategy used for optimizing the VQ in image synthesis \cite{zhang2023regularized} which used KL divergence to evaluate the discrepancy between the two distributions. 
The cross-entropy loss choice ensures more robust codebook optimization and mitigates issues related to unselected codewords during the early training phases. Specifically, KL divergence, defined as $\mathbb{D}_{KL}(P_{post}||P_{prior})=-\sum_{m=1}^M\sum_{k=1}^Kf_k^m\log\frac{f_k^m}{1/K}$, is prone to numerical instability due to the initially sparse utilization of codes, where $f_k^m = 0$ frequently occurs as some codes remain unselected, leading to a halt in the optimization process. In contrast, cross-entropy effectively addresses this issue, ensuring stable and reliable training.

\subsection{Inter-Codebook Optimization}
When viewed as a cohesive unit, RVQ is a residual structure composed of multiple stacked VQs. 
As the depth of the RVQ increases, redundancy may arise, characterized by overlapping focus on similar signal features across different layers within the RVQ structure. Prior studies, such as SpeechTokenizer \cite{zhang2024speechtokenizer}, have observed that the initial layers of RVQ primarily capture content-related features, while the later layers tend to encode acoustic information. This indicates a potential for redundancy, as multiple layers might concentrate on similar acoustic characteristics, such as speaker traits, rather than diversifying their representations. 
To address this issue and enhance the diversity of the quantized features, we propose introducing a loss function between adjacent VQs within the RVQ structure. This loss function is explicitly designed to encourage each VQ to focus on distinct aspects of the vectors being quantized, namely the encoded speech features, thereby reducing redundancy and improving the efficiency of information representation within each VQ codebook. This approach ensures that adjacent quantizers contribute unique and complementary information, ultimately improving the overall expressiveness and performance of the RVQ. 

We propose measuring the SSIM \cite{wang2004image} between the vectors quantified by the $m$-th VQ and the $(m+1)$-th VQ and minimizing their sum using the following loss function:
\begin{equation}
    \mathcal{L}_{SSIM}=\sum_{m=1}^{M-1} SSIM(\hat{\textbf{z}}^{m}, \hat{\textbf{z}}^{m+1}),
\end{equation}
where $\hat{\textbf{z}}^{m}$ represents the quantized output of the $m$-th VQ, and $M$ denotes the total number of VQs in the RVQ. The $SSIM(\cdot,\cdot)$ is calculated as follows:
\begin{equation}
    SSIM(\textbf{x},\textbf{y}) = \frac{(2\mu_\textbf{x}\mu_\textbf{y}+C_1)(2\sigma_{\textbf{x}\textbf{y}}+C_2)}{(\mu_\textbf{x}^2+\mu_\textbf{y}^2+C_1)(\sigma_\textbf{x}^2+\sigma_\textbf{y}^2+C_2)},
\end{equation}
where $\mu_\textbf{x}, \mu_\textbf{y}$ and $\sigma_\textbf{x}^2, \sigma_\textbf{y}^2$ are the means and variances of $\textbf{x}$ and $\textbf{y}$, respectively. $\sigma_{\textbf{x}\textbf{y}}$ is the covariance of $\textbf{x}$ and $\textbf{y}$, and $C_1, C_2$ are constants. 
SSIM is more robust than mean square error (MSE) in capturing perceptual differences and local structure, making it especially suitable for regulating latent feature similarity. 
In practice, we apply the SSIM loss only between adjacent VQ layers. This is based on the recursive structure of RVQ, where each quantizer encodes the residual of the previous one. Since adjacent layers are directly related, this localized regularization effectively encourages feature diversity while keeping the computational cost manageable. Extending the SSIM loss to all quantizer pairs introduces quadratic computational overhead with potentially limited additional benefit. Empirically, we also observed that applying SSIM across all VQ pairs in APCodec led to unstable gradient dynamics and hindered convergence during training.
By utilizing the above loss function, the overall discriminative ability of the RVQ is improved. This enhancement allows the encoder to capture more distinct signal features, thereby increasing the efficiency and effectiveness of the encoding process. 

\subsection{Overall Training Criterion}
Considering the original loss function of the neural audio codec as $\mathcal{L}_{codec}$, after applying our proposed ERVQ strategy, the total training loss function becomes
\begin{equation}
    \mathcal{L} = \mathcal{L}_{codec} + \alpha \mathcal{L}_{balancing} + \beta\mathcal{L}_{SSIM},
\end{equation}
where $\alpha$ and $\beta$ are the weights for the code balancing loss and the inter-codebook SSIM loss. Additionally, the traditional VQ module is replaced by an online clustering VQ module while applying the proposed ERVQ strategy. Therefore, our proposed ERVQ strategy only affects the training process of the codec, does not introduce any additional parameters to the model nor decrease its inference efficiency, and can be easily integrated into any neural audio codec structure which employs RVQ and its variants. 

\section{Experiments: Audio Coding}
\label{sec: Experiments: Audio Coding}
The most crucial application of audio codecs is audio communication and transmission. Therefore, we first evaluated the impact of the proposed ERVQ strategy on improving the performance of audio codecs in the audio coding field. 

\subsection{Experimental Details}

\subsubsection{Implementation}
To verify the effectiveness and generalization ability of our proposed ERVQ strategy, we implemented it on several existing audio codecs with various architectures and bitrates to compress audio at different sampling rates. Specifically, we applied the proposed ERVQ strategy to five open-source audio codecs. For non-streamable codecs, we used HiFi-Codec\footnote{\url{https://github.com/yangdongchao/AcademiCodec}.} \cite{yang2023hifi}, DAC\footnote{\url{https://github.com/descriptinc/descript-audio-codec}.} \cite{kumar2024high}, and APCodec\footnote{\url{https://github.com/YangAi520/APCodec}.} \cite{ai2024apcodec}. For streamable codecs, we utilized Encodec\footnotemark[2] \cite{defossez2023high} and APCodec-S\footnotemark[4] \cite{ai2024apcodec}. All hyper-parameters were set following their original codes, except that the standard VQ modules were replaced with the online clustering VQ modules in RVQ, and the code balancing loss and the SSIM loss were added to the final codec loss function. In all experiments, we set $\gamma = 0.999$ and $\epsilon = 1\times 10^{-3}$ in equation (2) and (3) following the configuration used in \cite{zheng2023online}. These values have been shown to provide stable training dynamics in image reconstruction task and worked well in our multi-codebook setting without further tuning.

Notably, when applying the ERVQ strategy to Encodec, we replaced the original codebook reset strategy with the online clustering strategy. For HiFi-Codec using GRVQ, the ERVQ strategy was applied separately to each group within the GRVQ, where the code balancing loss and SSIM loss obtained from each group were combined for gradient backpropagation. For DAC and APCodec, we retained the original factorized and L2 normalized codes strategies, integrating the ERVQ strategy directly with them. These settings show that our method can be easily applied to any existing architecture. All models were trained from scratch on our constructed datasets for fair comparison.

\subsubsection{Datasets} 
For the main experiments, we used a subset of the VCTK-0.92 corpus \cite{vctk} following APCodec \cite{ai2024apcodec}, which contains approximately 43 hours of 48 kHz speech recordings from 108 speakers. We selected 40,936 utterances from 100 speakers for the training set, while the test set comprised 2,937 utterances from the remaining 8 unseen speakers. Both the original 48 kHz waveforms and downsampled waveforms at 24 kHz and 16 kHz were used in the experiments to train all five codecs. 
We additionally validated the generalizability of ERVQ on a large-scale LibriTTS dataset \cite{zen2019libritts} under a 16kHz / 4kbps configuration with Encodec for experimental efficency. We followed the official configuration \cite{zen2019libritts}, using train-clean-100 and train-clean-360 as the training set, and dev-clean and test-clean as the validation and test sets, respectively. The model was trained on LibriTTS training set from scratch.
To evaluate the generalization performance of codecs trained with the ERVQ strategy on non-speech audio datasets, we conducted additional experiments using two publicly available datasets: FSD50K \cite{fonseca2021fsd50k} and Opencpop \cite{wang22b_interspeech}.
FSD50K \cite{fonseca2021fsd50k} is a human-labeled sound event dataset with a sampling rate of 44.1 kHz, comprising approximately 84 hours of recordings. 
Following the set partitioning in \cite{ai2024apcodec}, 40,966 utterances were used for training, and 10,231 utterances were designated for testing.
Opencpop \cite{wang22b_interspeech} is a high-quality Mandarin singing corpus, also sampled at 44.1 kHz, with a total duration of approximately 5.2 hours. 
Utilizing the officially pre-trimmed data, we selected 3,367 utterances as the training set and the remaining 389 utterances as the test set. 
For training, the FSD50K and Opencpop datasets were resampled to different sampling rates for different codecs. We initially trained all codec models on the VCTK dataset, then fine-tuned them separately on the Opencpop and FSD50K datasets to obtain models tailored for each dataset.

\begin{table*}[]
\centering
\caption{
Objective evaluation results of various neural audio codecs with and without the proposed ERVQ strategy on speech test set. Numbers in parentheses represent p-values from paired t-tests. All results, except for the last row, are obtained on the VCTK test set; the last row corresponds to the LibriTTS test set.}
\label{Table: Main Results}
\begin{tabular}{c|c|c|c|c|c|ccc}
\toprule
Model & Dataset & Streamable           & Sampling Rate           & Bitrate                 & ERVQ                        & ViSQOL($\uparrow$)  & STOI($\uparrow$)    & LSD($\downarrow$)   \\ \hline
\multirow{3}{*}{HiFi-Codec} & \multirow{18}{*}{VCTK} & \multirow{9}{*}{No}  & \multirow{3}{*}{16 kHz} & \multirow{3}{*}{2 kbps} & \XSolidBrush & 3.996               & 0.851               & 0.996               \\
                            &                      &                         &          &               & \Checkmark   & \textbf{4.235}      & \textbf{0.880}      & \textbf{0.975}      \\
                            &                      &                         &           &              & & (\textless{}1e-308) & (\textless{}1e-308) & (\textless{}1e-308) \\ \cline{1-1} \cline{4-9} 
\multirow{3}{*}{DAC}        &         &             & \multirow{3}{*}{24 kHz} & \multirow{3}{*}{3 kbps} & \XSolidBrush & 4.282               & 0.887               & 0.883               \\
                            &                      &                         &            &             & \Checkmark   & \textbf{4.286}      & \textbf{0.893}      & \textbf{0.880}      \\
                            &                      &                         &             &            & & (8.35e-05)          & (4.51e-21)          & (3.50e-104)         \\ \cline{1-1} \cline{4-9} 
\multirow{3}{*}{APCodec}    &         &             & \multirow{3}{*}{48 kHz} & \multirow{3}{*}{6 kbps} & \XSolidBrush & 4.070               & 0.875               & 0.818               \\
                            &                      &                         &              &           & \Checkmark   & \textbf{4.204}      & \textbf{0.888}      & \textbf{0.809}      \\
                            &                      &                         &               &          & & (\textless{}1e-308) & (\textless{}1e-308) & (\textless{}1e-308) \\ \cline{1-1} \cline{3-9} 
\multirow{3}{*}{Encodec} &   & \multirow{12}{*}{Yes} & \multirow{3}{*}{24 kHz} & \multirow{3}{*}{4 kbps} & \XSolidBrush & 4.384              & 0.851            & 1.072\\
                         &                      &                         &                      &   & \Checkmark   & \textbf{4.401}& \textbf{0.853}& \textbf{1.062}\\
                         &                      &                         &                       &  &                             & (9.52e-8)          & (3.67e-4)        & (\textless{}1e-308) \\ \cline{1-1} \cline{4-9} 
\multirow{3}{*}{Encodec}    &           &        & \multirow{3}{*}{24 kHz} & \multirow{3}{*}{6 kbps} & \XSolidBrush & 4.471               & 0.937               & 0.858               \\
                            &                      &                         &                  &       & \Checkmark   & \textbf{4.506}      & \textbf{0.939}      & \textbf{0.824}      \\
                            &                      &                         &                   &      & & (2.10e-306)         & 4.31e-98            & (\textless{}1e-308) \\ \cline{1-1} \cline{4-9} 
\multirow{3}{*}{APCodec-S}  &         &             & \multirow{3}{*}{48 kHz} & \multirow{3}{*}{6 kbps} & \XSolidBrush & 3.927               & 0.865               & 0.835               \\
                            &                      &                         &                    &     & \Checkmark   & \textbf{3.976}      & \textbf{0.877}      & \textbf{0.829}      \\
                            &                      &                         &                     &    & & (4.60e-124)         & (\textless{}1e-308) & (\textless{}1e-308) \\ \cline{1-2} \cline{4-9} 
\multirow{3}{*}{Encodec} & \multirow{3}{*}{LibriTTS} &     & \multirow{3}{*}{16 kHz} & \multirow{3}{*}{4 kbps} & \XSolidBrush & 4.473               & 0.940               & 0.855\\
                         &                      &                         &                        & & \Checkmark   & \textbf{4.536}& \textbf{0.945}& \textbf{0.847}\\
                         &                      &                         &                         & &                             & (\textless{}1e-308) & (\textless{}1e-308) & (\textless{}1e-308) \\
\toprule
\end{tabular}
\end{table*}


\begin{table*}[]
\centering
\caption{Objective evaluation results of various neural audio codecs with and without the proposed ERVQ strategy on FSD50K and Opencpop test set. Numbers in parentheses indicate p-values in paired t-tests.}
\label{Table: Main Results non-speech}
\begin{tabular}{c|c|c|c|cc|cc}
\toprule
\multirow{2}{*}{Model}      & \multirow{2}{*}{Sampling Rate} & \multirow{2}{*}{Bitrate} & \multirow{2}{*}{ERVQ}       & \multicolumn{2}{c|}{FSD50K}                                                 & \multicolumn{2}{c}{Opencpop}    \\ \cline{5-8} 
                            &                                &                          &                             & ViSQOL                               & LSD                                  & ViSQOL         & LSD            \\ \hline
\multirow{3}{*}{HiFi-Codec} & \multirow{3}{*}{16kHz}         & \multirow{3}{*}{2kbps}   & \XSolidBrush & 2.434                                & 1.187                                & 2.998          & 1.412          \\
                            &                                &                          & \Checkmark   & \textbf{2.669}                       & \textbf{1.163}                       & \textbf{3.245} & \textbf{1.390} \\
                            &                                &                          &                             & (\textless{}1e-308) & (1.06e-12)                           & (1.58e-90)     & (3.33e-19)     \\ \hline
\multirow{3}{*}{DAC}& \multirow{3}{*}{24kHz}& \multirow{3}{*}{3kbps}& \XSolidBrush & 4.225& 1.024& 4.347& 1.041\\
                            &                                &                          & \Checkmark   & \textbf{4.231}& \textbf{0.983}& \textbf{4.405}& \textbf{1.026}\\
                            &                                &                          &                             & (3.45e-2)& (\textless{}1e-308)& (3.54e-540)& (4.09e-2)\\ \hline
\multirow{3}{*}{Encodec}& \multirow{3}{*}{24kHz}& \multirow{3}{*}{6kbps}   & \XSolidBrush & 4.331               
& 1.090& 4.510               
& 1.219\\
                            &                                &                          & \Checkmark   & \textbf{4.347}& \textbf{1.078}& \textbf{4.526}& \textbf{1.218}\\
                            &                                &                          &                             & (4.75e-6)& (5.51e-45)& (3.50e-2)           & (5.00e-2)\\ \hline
\multirow{3}{*}{APCodec}    & \multirow{3}{*}{48kHz}         & \multirow{3}{*}{6kbps}   & \XSolidBrush & 3.705                                & 0.949                                & 3.818          & 1.003          \\
                            &                                &                          & \Checkmark   & \textbf{4.054}                       & \textbf{0.852}                       & \textbf{4.302} & \textbf{0.863} \\
                            &                                &                          &                             & (\textless{}1e-308) & (\textless{}1e-308) & (3.44e-138)    & (8.35e-175)   \\
\toprule
\end{tabular}
\end{table*}


\subsubsection{Metrics} 
We comprehensively evaluated the performance of the decoded speech on VCTK test set through multiple objective metrics. The virtual speech quality objective listener (ViSQOL) tool\footnote{\url{https://github.com/google/visqol}.} \cite{hines2015visqol} was used to evaluate the overall quality of the decoded speech. ViSQOL provides a mean opinion score-listening quality objective (MOS-LQO), ranging from 1 to 4.75 for 48 kHz and 1 to 5 for 16 kHz sampling rates. It is important to note that ViSQOL supports only 48 kHz and 16 kHz sampling rates. To assess speech quality at a 24 kHz sampling rate, we upsampled both the decoded and reference speech to 48 kHz and then calculated the MOS-LQO using ViSQOL’s 48 kHz mode, following the method described in \cite{kumar2024high}. To quantify the decoded speech's intelligibility, we utilized the short-time objective intelligibility (STOI) \cite{taal2010short} metric. The commonly used log spectral distance (LSD) metric was also employed to assess the amplitude spectrum quality of the decoded speech.  
For the audio decoded on FSD50K and Opencpop test set, we only used ViSQOL and LSD as evaluation metrics.

We further conducted an ABX preference listening test on the Amazon Mechanical Turk platform\footnote{\url{https://www.mturk.com/}.} to subjectively evaluate the audio codecs before and after employing the ERVQ strategy if they have subtle differences in objective metrics. Specifically,  in each ABX test, 20 utterances were randomly selected from the test set decoded by the original and ERVQ-enhanced neural audio codecs and evaluated by at least 20 native English listeners. The listeners were asked to judge which utterance in each pair had better speech quality or whether there was no preference. 

\begin{figure}[t]
\centering
\includegraphics[width=0.50\textwidth]{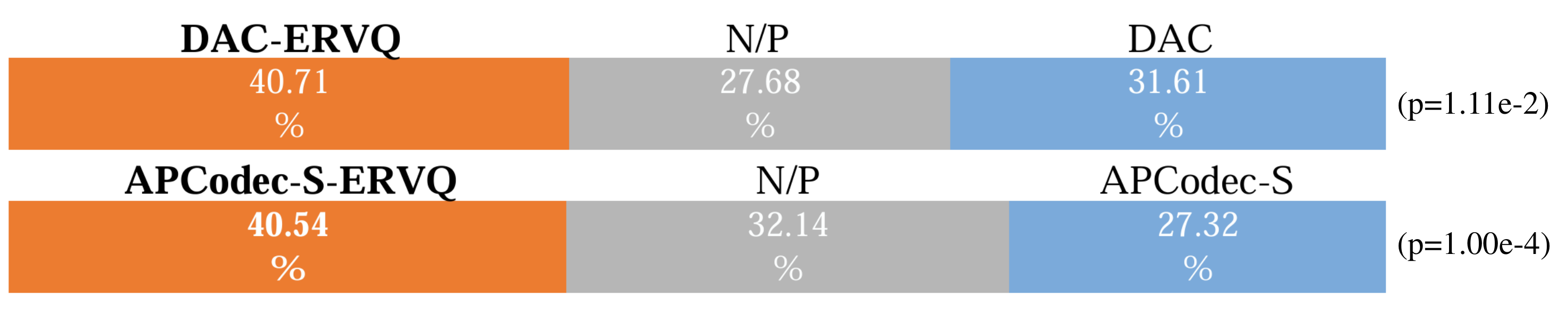} 
\caption{ABX preference listening tests results on DAC \cite{kumar2024high} at a sampling rate of 24 kHz and a bitrate of 3 kbps and APCodec-S \cite{ai2024apcodec} at a sampling rate of 48 kHz and a bitrate of 6 kbps. }
\label{Figure: ABX}
\end{figure}

\subsection{Main Results}

The first seven rows of Table \ref{Table: Main Results} presents the objective evaluation results on constructed VCTK test set. Across various model structures and bitrates, neural audio codecs utilizing our proposed ERVQ strategy consistently outperformed all baselines in speech coding performance, with particularly notable improvements for HiFi-Codec, APCodec, and Encodec. 
In particular, the comparison between the fifth and sixth rows in Table \ref{Table: Main Results} shows that the proposed ERVQ strategy is not sensitive to the number of quantization stages or bitrate constraints.
As shown in the last row of Table \ref{Table: Main Results}, ERVQ continued to yield consistent improvements over the baseline on LibriTTS dataset, demonstrating its effectiveness across varying dataset scales.
A similar trend is also observed when evaluated on the FSD50K and Opencpop test sets, as shown in Table \ref{Table: Main Results non-speech}, further confirming the effectiveness of the ERVQ strategy in audio coding. 
The improvements are statistically significant in all metrics, with p-value $\textless{} 0.05$ in paired t-tests.

For the results of DAC and APCodec-S on VCTK test set, we conducted ABX preference listening tests due to subtle differences in their objective metrics. The results demonstrated in Figure \ref{Figure: ABX} showed that 40.71\% of participants preferred the speech decoded by DAC enhanced with ERVQ, compared to 31.61\% who preferred the original DAC results. For APCodec-S, these proportions were 40.54\% and 27.32\%, respectively. Paired t-tests revealed p-values less than 0.05, indicating the statistical significance of these subjective preferences.

These results demonstrate that the proposed ERVQ strategy can significantly enhance the performance of various neural audio codecs, underscoring its effectiveness in optimizing the RVQ framework. 

\subsection{Component Analysis}
We conducted a series of experiments to analyze the effectiveness of each part of the proposed ERVQ strategy. For the sake of experimental efficiency, the main component analysis experiments were performed on APCodec at a 48 kHz sampling rate and a bitrate of 6 kbps on VCTK dataset. Results are reported in row 2-7 of Table \ref{Table: Ablation}. To further illustrate  the advantages of the online clustering strategy over the codebook reset strategy \cite{dhariwal2020jukebox} in neural audio codecs, we performed  analytical experiments on Encodec \cite{defossez2023high} at a 24 kHz sampling rate and a bitrate of 6 kbps. Specifically, we added code balancing loss and SSIM loss to the original Encodec for training without employing online clustering VQs. Results are shown in the last three rows of Table \ref{Table: Ablation}.

\begin{table*}[]
\centering
\caption{Component analysis for the proposed ERVQ strategy. Experiments were conducted using APCodec at a 48 kHz sampling rate and a bitrate of 6 kbps and Encodec at a 24kHz sampling rate and a bitrate of 6kbps. ``OC" and ``CB" represents the online clustering strategy and the code balancing loss for short, respectively. The content in brackets indicates different anchor sampling methods.}
\label{Table: Ablation}
\begin{tabular}{c|ccccc|ccc}
\toprule
Model                    & +OC                         & +CB                         & +SSIM                       & +OC(Random)                 & +OC(Closest)                & ViSQOL($\uparrow$) & STOI($\uparrow$) & LSD($\downarrow$) \\ \hline
\multirow{6}{*}{APCodec} & \XSolidBrush & \XSolidBrush & \XSolidBrush & \XSolidBrush & \XSolidBrush & 4.070              & 0.875            & 0.818             \\
                         & \Checkmark   & \XSolidBrush & \XSolidBrush & \XSolidBrush & \XSolidBrush & 4.172              & 0.882            & 0.815             \\
                         & \Checkmark   & \Checkmark   & \XSolidBrush & \XSolidBrush & \XSolidBrush & 4.146              & 0.887            & 0.807             \\
                         & \Checkmark   & \Checkmark   & \Checkmark   & \XSolidBrush & \XSolidBrush & 4.204              & 0.888            & 0.809             \\
                         & \XSolidBrush & \XSolidBrush & \XSolidBrush & \Checkmark   & \XSolidBrush & 4.105              & 0.881            & 0.816             \\
                         & \XSolidBrush & \XSolidBrush & \XSolidBrush & \XSolidBrush & \Checkmark   & 4.131              & 0.876            & 0.813             \\ \hline
\multirow{3}{*}{Encodec} & \XSolidBrush & \XSolidBrush & \XSolidBrush & \XSolidBrush & \XSolidBrush & 4.471              & 0.937            & 0.858             \\
                         & \XSolidBrush & \Checkmark   & \Checkmark   & \XSolidBrush & \XSolidBrush & 4.493              & 0.931            & 0.832             \\
                         & \Checkmark   & \Checkmark   & \Checkmark   & \XSolidBrush & \XSolidBrush & 4.506              & 0.939            & 0.824             \\ 
\toprule
\end{tabular}
\end{table*}

\begin{figure*}[t]
\centering
\includegraphics[width=0.95\textwidth]{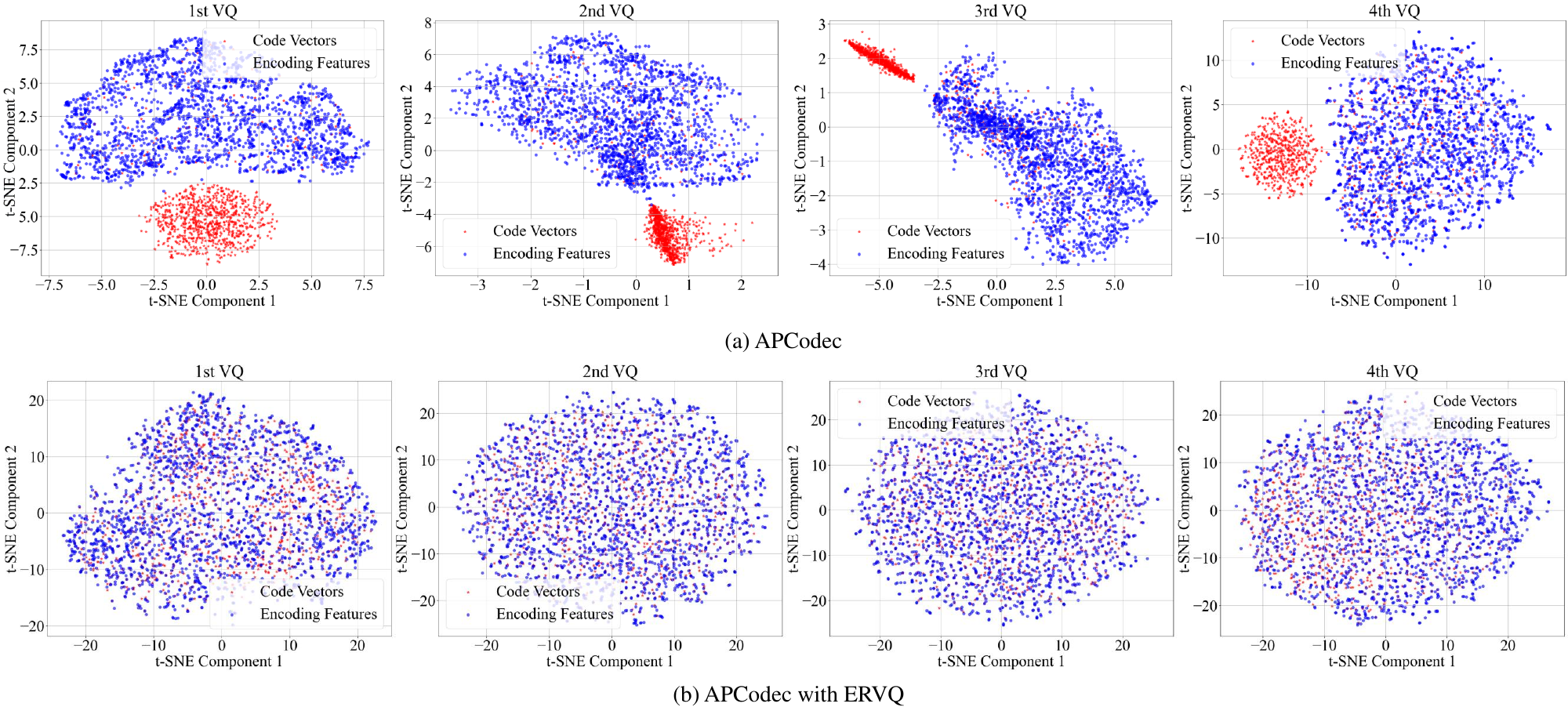} 
\caption{T-SNE visualization results of each VQ's encoding features and codebooks in APCodec \cite{ai2024apcodec} at a sampling rate of 48 kHz and a bitrate of 6 kbps. The blue circle points and red star points represent the features to be quantized from 10 utterances in the test set and codec vectors in the codebook, respectively.}
\label{Figure: Codebook Analysis}
\end{figure*}

\subsubsection{Core factors}

The effectiveness of the core components in the proposed ERVQ strategy is demonstrated in row 2-5 and 8-10 of Table \ref{Table: Ablation}. A comparison among the first three rows shows that while the online clustering strategy significantly enhances system performance, incorporating the code balancing loss further improves intelligibility and reduces amplitude distortion. Although the ViSQOL score slightly decreases when the code balancing loss is added, the improvements in STOI and LSD indicate enhanced intelligibility and spectral fidelity. This suggests that the CB loss promotes more effective code usage, even if its effect on perceptual quality is less pronounced. Additionally, introducing the SSIM loss markedly boosts both the overall quality and intelligibility of the speech. 
Furthermore, inferior results were observed on Encodec when only the code balancing loss and SSIM loss were applied, highlighting the superiority of the online clustering strategy over the codebook reset strategy. These findings collectively validate the effectiveness of each component in the ERVQ framework.

\subsubsection{Anchor Sampling Methods}
An evaluation of various anchor sampling methods is reported in row 3 and 6-7 in Table \ref{Table: Ablation}. Two additional anchor sampling methods were explored besides the aforementioned probabilistic random sampling method: closest and random sampling. For the closest sampling method, the anchor is selected by inversely looking up the closest features of each entry, i.e., $\mathop{\arg\min}\limits_{\textbf{z}_i}|\textbf{z}_i^{(t)}-\textbf{e}_k^{(t)}|^2$. For the random sampling method, the anchors are randomly sampled from the encoded features. In this ablation,  we removed the proposed code balancing and SSIM loss function and retained only the online clustering strategy.
In contrast to the findings in \cite{zheng2023online}, our results indicate that the choice of anchor sampling methods could significantly impact the audio codec's overall performance. 
Paired t-tests on ViSQOL scores of different anchor sampling method indicate that the differences between the probabilistic random sampling method and both the random and closest methods are statistically significant ($p < 0.05$).
It reveals that the probabilistic random sampling method equipped the enhanced RVQ framework with the highest decoded speech quality. Consequently, we adopted probabilistic random sampling for all other experiments as the anchor sampling method.

\subsection{Codebook Analysis}

\begin{table*}[]
\centering
\caption{Analysis of each codebook within the VQ from the RVQ in APCodec at a sampling rate of 48 kHz and a bitrate of 6 kbps. Each VQ has a codebook size of 1024. \# indicates the number of the VQ in the RVQ.}
\label{Table: Codebook Analysis}
\begin{tabular}{c|cccc|cccc|c}
\toprule
\multirow{2}{*}{ERVQ}& \multicolumn{4}{c|}{Utilization Rate/\%($\uparrow$)}                  & \multicolumn{4}{c|}{Perplexity($\uparrow$)}                           & \multirow{2}{*}{BE($\uparrow$)} \\ \cline{2-9}
                        & \#1       & \#2       & \#3       & \#4       & \#1       & \#2       & \#3       & \#4       &                                     \\ \hline
\XSolidBrush& 14.7         & 16.3         & 25.5         & 41.2         & 102          & 157          & 256          & 401          & 0.766                                        \\
\Checkmark& \textbf{100} & \textbf{100} & \textbf{100} & \textbf{100} & \textbf{653} & \textbf{920} & \textbf{973} & \textbf{962} & \textbf{0.976}                      \\
\toprule
\end{tabular}
\end{table*}

To further demonstrate the effectiveness of our proposed ERVQ strategy, we analyzed the usage of the VQ codebooks within the RVQ framework of neural audio codecs. For experimental efficiency, we conducted the analysis on the VCTK test set using two representative codecs, i.e., HiFi-Codec and APCodec. In APCodec, which operates at a sampling rate of 48 kHz and a bitrate of 6 kbps, the RVQ consists of 4 VQs, each with a codebook size of 1024. The GRVQ in HiFi-Codec, operating at a sampling rate of 16 kHz and a bitrate of 2 kbps, comprises 2 groups of RVQ, each containing 2 VQs with a codebook size of 1024. 
To evaluate the usage of these codebooks, we measured their utilization rate, perplexity, and bit efficiency \cite{kumar2024high} for each VQ on the VCTK test set. 
\begin{itemize}
    \item \textbf{Utilization Rate}: The codebook utilization rate is calculated as $\frac{\sum_k\mathbf{1}_{[n_k>0]}}{K}$, where $n_k$ represents the number of features quantized to the $k$-th code, and $\mathbf{1}_{[n_k>0]}$ is an indicator function that equals 1 if codeword $k$ is used at least once and 0 otherwise. Ideally, a codebook without codebook collapse issue should have a utilization rate as close to 100\% as possible. 
    \item \textbf{Perplexity}: The perplexity of the codebook is calculated as $2^{-\sum_kp_k\log p_k}$, where $p_k=\frac{n_k}{N}$ represents the probability (frequency) of the features quantized to the $k$-th code, and $N$ is the total number of the features in the test set. Perplexity reflects the average uncertainty of each code. A higher perplexity suggests a more balanced utilization of the codebook, whereas a lower perplexity may indicate the presence of codebook collapse issues. 
    \item \textbf{Bitrate Efficiency} (\textbf{BE}) \cite{kumar2024high}: The bitrate efficiency is calculated as the sum of the entropy (in bits) of each codebook when applied on the test set divided by the number of bits across all codebooks, i.e., $\frac{-\sum_{m=1}^M\sum_{k=1}^Kp_k^m\log p_k^m}{M\log K}$. For efficient bitrate utilization this should tend to 100\% and lower percentages indicate that the bitrate is being underutilized. 
\end{itemize}

\begin{figure*}[t]
\centering
\includegraphics[width=0.95\textwidth]{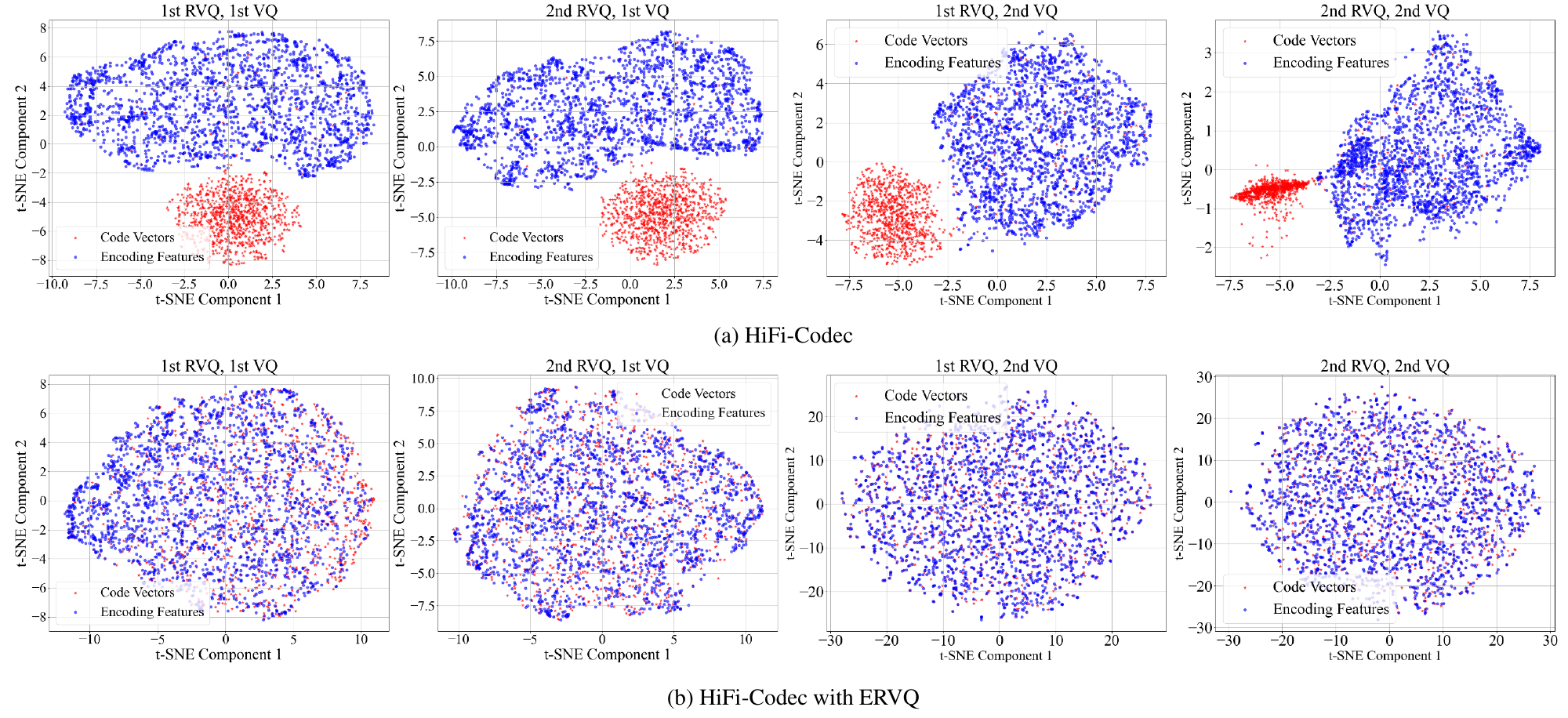} 
\caption{T-SNE visualization results of each VQ's encoding features and codebooks in HiFi-Codec \cite{defossez2023high} at a sampling rate of 16 kHz and a bitrate of 2 kbps. The blue circle points and red star points represent the features to be quantized from 10 utterances in the test set and codec vectors in the codebook, respectively.}
\label{Figure: Codebook Analysis HiFiCodec}
\end{figure*}

The codebook analysis results of APCodec are displayed in Table \ref{Table: Codebook Analysis}. Additionally, Figure \ref{Figure: Codebook Analysis} shows the t-SNE visualization results of the encoding features to be quantized from 10 utterances in the test set and the code vectors in the codebook of each VQ from APCodec, both before and after using the ERVQ strategy. The experimental results reveal that our proposed ERVQ strategy achieved 100\% codebook usage for each VQ in the RVQ from APCodec, significantly improving codebook perplexity and bit efficiency, with bit efficiency increasing by more than 27\%. The visualization results also show that most code vectors in the codebook were inactive and unused before the ERVQ strategy was applied. In contrast, after implementing the ERVQ strategy, the codebook collapse issue was resolved. 
A similar trend is observed in HiFi-Codec, which employs the GRVQ structure, as shown in Table \ref{Table: Codebook Analysis HiFiCodec} and Figure \ref{Figure: Codebook Analysis HiFiCodec}. The ERVQ strategy significantly improved code utilization across all VQ codebooks, particularly for the first VQ in each RVQ group, where utilization rates increased from 6.05\% to 98.6\% and 5.57\% to 99.8\%. These findings confirm the effectiveness of the ERVQ strategy in addressing codebook collapse and enhancing codebook expressiveness in neural audio codecs. 

An ineresting finding is that, although the proposed method significantly increases codebook utilization, the corresponding gains in audio reconstruction metrics are relatively modest. This phenomenon reflects the fact that not all codewords contribute equally to perceived quality, and speech inherently contains structured redundancy. 
We consider this an important direction for future investigation, as perfectly uniform code usage may not be optimal, particularly in the presence of context-dependent or temporally specific patterns.

\begin{table*}[]
\centering
\caption{Analysis of each codebook within the VQ from the GRVQ in HiFi-Codec at a sampling rate of 16 kHz and a bitrate of 2 kbps. Each VQ has a codebook size of 1024. \#m\_k indicates the k-th VQ in the m-th RVQ group.}
\label{Table: Codebook Analysis HiFiCodec}
\begin{tabular}{c|cccc|cccc|c}
\toprule
\multirow{2}{*}{ERVQ}       & \multicolumn{4}{c|}{Utilization Rate($\uparrow$)}            & \multicolumn{4}{c|}{Perplexity($\uparrow$)}               & \multirow{2}{*}{BE($\uparrow$)} \\ \cline{2-9}
                            & \#1\_1        & \#1\_2       & \#2\_1        & \#2\_2        & \#1\_1       & \#1\_2       & \#2\_1       & \#2\_2       &                                 \\ \hline
\XSolidBrush & 6.05          & 14.3         & 5.57          & 14.1          & 24.1         & 111          & 28.7         & 110          & 0.575                           \\
\Checkmark   & \textbf{98.6} & \textbf{100} & \textbf{99.8} & \textbf{99.8} & \textbf{364} & \textbf{703} & \textbf{480} & \textbf{600} & \textbf{0.908}                 \\
\toprule
\end{tabular}
\end{table*}

\section{Experiments: LLM Applications}
\label{sec: Experiments: LLM Applications}

One of the promising applications of audio codecs is the incorporation into unified speech-and-text LLMs. To evaluate whether an ERVQ-enhanced codec can improve speech-and-text LLMs, we conducted a series of experiments. Specifically, we used FunCodec \cite{du2024funcodec} as the audio codec to provide intermediate speech representations and LauraGPT \cite{chen2023lauragpt} as the corresponding speech-and-text LLM, as they have official open-source training recipes\footnote{\url{https://github.com/modelscope/FunCodec}.}. Experimental details and results are described as follows.

\subsection{Constructing FunCodec with ERVQ Strategy}

\subsubsection{Implementation Details}
We adopt FunCodec \cite{du2024funcodec} for our LLM experiments to match the original architecture of LauraGPT \cite{chen2023lauragpt}. This ensures compatibility and allows for a fair comparison when assessing the impact of ERVQ within the LauraGPT framework.
FunCodec \cite{du2024funcodec} is used as the audio tokenizer to extract discrete speech representations. It shares a similar architecture to Encodec \cite{defossez2023high}, comprising an encoder-RVQ-decoder structure. Its key improvement over Encodec is the incorporation of reconstruction losses in the amplitude spectrum domain. In our experiments, we configured the strides in the encoder's convolutional blocks to $[8,5,4,2,2]$, following the setup in LauraGPT \cite{chen2023lauragpt}. We utilized 16 VQs, each with a codebook size of 1024 within the RVQ framework, resulting in a bitrate of 4 kbps for the sake of experimental efficiency. All other hyperparameters were set according to the original code. All the models were training from scratch on our constructed dataset.

\subsubsection{Datasets}
The commonly used LibriTTS corpus \cite{zen19_interspeech}, consisting of 585 hours of English speech, was downsampled to 16 kHz and employed to train and evaluate FunCodec. The dataset was partitioned into training, validation and test sets according to the official guidelines. 

\subsubsection{Experimental Results}

The experimental results of FunCodec, trained with and without the proposed ERVQ strategy, are presented in Table \ref{Table: FunCodec}. The results indicate that after applying the ERVQ strategy, FunCodec achieved statistically significant improvements across all evaluation metrics (p-value $<$ 0.05). These findings further validate the effectiveness of the ERVQ strategy in enhancing the performance of audio codecs.

\begin{table}[]
\centering
\caption{Objective evaluation results of decoded speech from FunCodec with and without the proposed ERVQ strategy. Numbers in parentheses indicate p-values in paired t-tests.}
\label{Table: FunCodec}
\begin{tabular}{c|ccc}
\toprule
ERVQ                        & ViSQOL($\uparrow$) & STOI($\uparrow$) & LSD($\downarrow$) \\ \hline
\XSolidBrush & 4.517              & 0.966            & 1.022             \\
\Checkmark   & \textbf{4.524}     & \textbf{0.967}   & \textbf{1.011}    \\
             & (2.73e-30)     & (1.10e-131)   & (2.00e-72)    \\
\toprule
\end{tabular}
\end{table}

\subsection{Applying the Enhanced FunCodec for LauraGPT}

\subsubsection{Implementation Details}
LauraGPT \cite{chen2023lauragpt} is a versatile unified speech-and-text LLM based on a decoder-only transformer architecture. It includes three main components: a generative pretrained transformer (GPT) backbone, an audio encoder, and a codec vocoder. LauraGPT employs continuous features from the audio encoder to represent input audio and discrete codec tokens from FunCodec for audio outputs. 
When incorporating FunCodec for LauraGPT, the encoder and the first VQ in RVQ were used as the audio tokenizer, with the outputs of the first quantizer serving as the audio tokens. These tokens are jointly autoregressively modeled with text features. For audio generation, LauraGPT utilizes a one-step codec vocoder, which includes a transformer-based predictor trained to estimate the sum of all codec token groups by minimizing reconstruction losses. We trained two versions of LauraGPT on the 16 kHz LibriTTS dataset: LauraGPT with the original FunCodec and LauraGPT with FunCodec enhanced by ERVQ. All training hyperparameters were set according to the original code. All the models were training from scratch on our constructed dataset.

\subsubsection{Metrics}
We evaluated the performance of LauraGPT on downstream zero-shot TTS tasks. For objective evaluation, we utilized character error rate (CER) and word error rate (WER) to assess content consistency. We utilized the ASR paraformer model \cite{gao22b_interspeech} provided in FunASR\footnote{\url{https://github.com/modelscope/FunASR}.} \cite{gao2023funasr} to transcribe the synthesized speech.  For reference, it achieved a 2.580\% CER and 7.107\% WER on the clean test set. Speaker similarity (SS) was assessed by calculating the cosine similarity between the speaker embeddings extracted from Resemblyzer\footnote{\url{https://github.com/resemble-ai/Resemblyzer}.} \cite{wan2018generalized}, a pre-trained speaker verification model for both the prompt speech and the synthesized speech using the Amphion toolkit\footnote{\url{https://github.com/open-mmlab/Amphion}.} \cite{zhang2023amphion}. Additionally, the naturalness of the synthesized speech was evaluated using UTMOS\footnote{\url{https://github.com/sarulab-speech/UTMOS22}.} \cite{saeki22c_interspeech}, a non-intrusive pre-trained scoring network. For subjective evaluation, we conducted a series of listening tests to measure the naturalness mean opinion scores (MOS) of synthesized speech, and the speaker similarity MOS (SMOS) of synthesized speech with the prompt audio. Specifically, 30 English speakers were recruited on Amazon’s Mechanical Turk\footnotemark[6] and were asked to give a 5-point score (1-very poor, 2-poor, 3-fair, 4-good, 5-excellent) for each utterance they listened to. 20 utterances from the test set were randomly selected for subjective evaluation.

\subsubsection{Experimental Results}
\begin{table*}[]
\centering
\caption{Evaluation results of LauraGPT on downstream zero-shot TTS task. ``ERVQ" represents whether LauraGPT is trained with an ERVQ-improved FunCodec or not. Numbers in parentheses indicate p-values in paired t-tests.}
\label{Table: LauraGPT}
\begin{tabular}{c|cccccc}
\toprule
ERVQ                        & CER($\downarrow$)            & WER($\downarrow$)            & SS($\uparrow$)& UTMOS($\uparrow$)          & MOS($\uparrow$)            & SMOS($\uparrow$)\\ \hline
\XSolidBrush & 3.393          & \textbf{7.470} & 0.784          & 4.383          & 3.753& 3.423\\
\Checkmark   & \textbf{3.125} & 7.716          & \textbf{0.797} & \textbf{4.397} & \textbf{3.940}& \textbf{3.668}\\
 & (3.91e-3)          & (1.08e-1) & (6.53e-34)          & (2.91e-6)          & (1.51e-4)& (3.09e-7) \\
\toprule
\end{tabular}
\end{table*}
The objective and subjective evaluation results of LauraGPT, trained with and without an ERVQ-improved FunCodec, are displayed in Table \ref{Table: LauraGPT}. The results indicate that training LauraGPT with codec tokens generated by the ERVQ-enhanced FunCodec significantly improves zero-shot TTS performance, particularly in terms of naturalness and speaker similarity. Notably, there is a notable 5\% improvement in MOS and a significant 7\% improvement in SMOS, with p-values of $1.15\times10^{-4}$ and $3.09\times10^{-7}$ in paired t-tests. We recommend readers refer to our demo page for audio samples. This enhancement is attributed to the ERVQ strategy, which enables the audio codec to learn and preserve more speech details, allowing LauraGPT to produce more expressive and emotionally rich speech. The results also demonstrate the effectiveness of the ERVQ strategy in enhancing the learning capability of the RVQ module within the audio codec, suggesting promising potential for practical applications.

\section{Conclusion}
\label{sec: Conclusion}

In this study, we introduced ERVQ, a comprehensive optimization strategy designed to enhance the RVQ framework used in neural audio codecs. Through intra-codebook and inter-codebook optimizations, ERVQ ensures more balanced and effective codebook utilization. Our extensive experiments on various audio codecs confirmed that ERVQ consistently improves the performance and quality of audio compression and reconstruction. Additionally, the integration of the ERVQ-enhanced audio codec into unified speech-and-text LLMs demonstrated significant enhancements on downstream zero-shot TTS tasks, underscoring its practical applicability. Our future work aims to optimize the RVQ structure within the audio codec to prevent excessive audio information from being quantized and stored in the first VQ, thereby enhancing the quantization capability of the audio codec. 
Extending ERVQ to scalar quantization frameworks can also be a promising future direction.

\bibliographystyle{IEEEtran}
\bibliography{mybib}


 





\end{document}